\documentclass[reprint,amsmath,amssymb,aps,superscriptaddress,twocolumn]{revtex4-2}
\usepackage{graphicx}
\usepackage{graphics}
\usepackage{dcolumn}
\usepackage{bm}
\usepackage{amsfonts}
\usepackage{amssymb}
\usepackage{xcolor}
\usepackage{multirow}
\usepackage{mathtools}

\definecolor{green}{rgb}{0,0.6,0.1}

\begin{document}

\title{Electron energy-loss spectrum and exciton band structure of ${\mathrm{WSe}}_{2}$ monolayer studied by 
{\it ab initio} Bethe-Salpeter equation calculations}

\author{Yun-Chen Shih}
\affiliation{Department of Physics and Center for Theoretical Physics, National Taiwan University, Taipei 10617, Taiwan\looseness=-1}

\author{Fredrik Andreas Nilsson}
\email{fanni@dtu.dk}
\affiliation{CAMD, Computational Atomic-Scale Materials Design, Department of Physics, Technical University of Denmark, 2800, Kgs. Lyngby, Denmark\looseness=-1}

\author{Guang-Yu Guo}
\email{gyguo@phys.ntu.edu.tw}
\affiliation{Department of Physics and Center for Theoretical Physics, National Taiwan University, Taipei 10617, Taiwan\looseness=-1}
\affiliation{Physics Division, National Center for Theoretical Sciences, Taipei 10617, Taiwan\looseness=-1}

\date{\today}

\begin{abstract}
Bounded excitons in transition metal dichalcogenides monolayers lead to numerous opto-electronic applications,
which require a detailed understanding of the exciton dynamics. 
The dynamical properties of excitons with finite momentum transfer $\textbf{Q}$ can be investigated experimentally 
using electron energy-loss (EEL) spectroscopy. The EEL spectrum depends on the response function of the material 
which in turn is determined by the exciton energies and eigenvectors in the exciton band structure.
In this work, we utilize {\it ab initio} density-functional theory plus Bethe-Salpeter equation (DFT+BSE) approach 
to explore the exciton band structure and also $\textbf{Q}$-resolved EEL spectrum in monolayer ${\mathrm{WSe}}_{2}$. 
In particular, we carefully examine the discrepancies and connections among the existing EEL spectrum formulas 
for quasi-two-dimensional (2D) systems, and establish a proper definition of the EEL spectrum, which is then used to 
calculate the EEL spectra of monolayer ${\mathrm{WSe}}_{2}$.
We find that remarkably, the dispersion of the calculated lowest-energy EELS peaks for the in-plane momentum transfer
follows almost precisely the non-parabolic upper band of the lowest bright A exciton, and also agrees well with the previous experiment.
Furthermore, we show that only the bright exciton with its electric dipole being parallel to the direction of the transfered momentum 
is excited, i.e., EEL spectroscopy selectively probes bright exciton bands.
This explains why only the upper band of the A exciton, which is a longitudinal exciton with
an in-plane dipole moment, was observed in the previous experiment.
Our findings will stimulate further EEL experiments to measure other branches of the exciton band structure, 
such as the parabolic lower band of the A exciton,
and hence will lead to a better understanding of the exciton dynamics in quasi-2D materials.

\end{abstract}

\maketitle
\section{INTRODUCTION}
Atomically thin quasi-two-dimensional (quasi-2D) materials, such as transition metal dichalcogenides (TMDs) monolayers, have received significant attention due to their exceptional physical properties. The combination of the strong spin-orbit coupling (SOC) effect  \cite{Zhu2011,Chhowalla2013,Zhao2013} and the broken inversion symmetry in the TMDs monolayers leads to the coupling between spin and valley degrees of freedom, and thus the valley dependent optical selection rules \cite{Sallen2012,DiXiao2012,Cao2012,Zeng2012,Mak2012}. Moreover, due to the weak dielectric screening \cite{Keldysh1979,Cudazzo2011,Falco2013,Latini2015} and the geometric confinement \cite{Yang1991}, the enhanced Coulomb interaction between the electrons and holes in the TMDs monolayers gives rise to the formation of the strongly bound excitons with exceptionally large binding energies \cite{Ramasubramaniam2012,Falco2013,Qiu2013,Chernikov2014,Mak2014} and huge oscillator strengths \cite{Mak2010,Tony2014}, leading to the promising opto-electronic and photonic applications \cite{Qing2012,Mak2016}.

Despite the violation of the momentum conservation rule in optical experiments, finite momentum excitons play the important roles as the de-excitation channels, significantly influence the photoluminescence and other time-dependent phenomena \cite{Zhang2015,Selig2018}. This has driven investigations into finite momentum excitons over the past few years \cite{Wu2015,Qiu2015,Cudazzo2016,Deilmann2019,Sauer2021,JinhuaHong2020}. For the TMDs monolayers, the excitation energies of the lowest bright excitons centered around the K and $\rm K'$ valleys are degenerate due to the time-reversal symmetry as $\textbf{Q} \rightarrow 0$ \cite{Qiu2015}. Remarkably, as demonstrated by the theoretical calculations \cite{Wu2015,Qiu2015,Cudazzo2016,Sauer2021,Qiu2021}, 
the dispersions of the lowest bright excitons split into two bands, with the lower band showing a parabolic dispersion, while the upper band exhibits a strong non-parabolic dispersion. The different dispersion behaviors can give rise to the different emission properties such as exciton lifetimes and angular-dependent emission behaviors \cite{Sauer2021}. 

Electron energy-loss (EEL) spectroscopy (EELS) is a powerful technique to probe plasmons \cite{Andersen2013,Liou2015,Nerl2017,JinhuaHong2022} 
and excitons with finite momenta \cite{Nerl2017,JinhuaHong2020,JinhuaHong2021}. 
In momentum-resolved EELS in transmission geometry the measured spectrum represents the differential cross-section of the inelastic electron scattering for a given momentum transfer $\textbf{Q}$ of the incident electrons \cite{Sturm1982}.
The EEL spectrum provides information of the dynamical response of the materials, such as the plasmonic excitations \cite{Yan2011,Andersen2013,Liou2015,JinhuaHong2022}, the strong excitonic effects \cite{Carsten2015,JinhuaHong2020,JinhuaHong2021,JinhuaHong2022}, and the localized plasmon-exciton coupling \cite{Nerl2017}. The theoretical studies \cite{Nerl2017,JinhuaHong2021,JinhuaHong2022} investigating the excitonic effects on the EEL spectra using Bethe-Salpeter equation (BSE) calculations primarily focus on the optical limit ($\textbf{Q} \rightarrow 0$). However, in practice, the experimental measurements can never reach exactly $\textbf{Q} \rightarrow 0$ \cite{Abajo2010,JinhuaHong2021}. Therefore, the theoretical analysis of the excitonic effects on the EEL spectra at non-vanishing $\textbf{Q}$ is required. In this work, we perform the state-of-the-art {\it ab initio} BSE calculations and present a comprehensive study of the significant excitonic effects on the EEL spectra in the finite $\textbf{Q}$ regime.

Conventionally, the EEL spectrum is defined as the inverse of the momentum- and energy-resolved macroscopic dielectric function 
($\varepsilon_{M}(\textbf{Q},\omega)$) as $L(\textbf{Q},\omega) = {\rm -Im[1/\varepsilon_{M}}(\textbf{Q},\omega)]$ \cite{Sturm1982,Sturm1993}.  
While this quantity is well-defined for three-dimensional (3D) bulk systems, for quasi-2D systems $\varepsilon_{M}$ is significantly influenced 
by the supercell volume adopted in periodic calculations. This has led to confusion about how to correctly calculate the EEL spectrum in 2D materials, 
with fundamentally different definitions used in different works \cite{Sturm1982,Sturm1993,Nerl2017,nazarov2015electronic,JinhuaHong2021}.
Therefore we begin by examining the most commonly used definitions of the EEL spectrum for quasi-2D systems, 
elaborate on the connections between these formulas, and provide a well-defined, easily implementable EEL spectrum formula for quasi-2D systems, 
suitable for the momentum and energy regimes studied in this work. We then focus on monolayer ${\mathrm{WSe}}_{2}$ and analyse its EEL spectrum
and exciton band structure, and also compare the calculated EEL spectrum with the recent experiment \cite{JinhuaHong2020}.
Similar to that reported by Qiu $et\,al.$ \cite{Qiu2015} for monolayer ${\mathrm{MoS}}_{2}$ and by Sauer $et\,al.$ \cite{Sauer2021} 
for the ${\mathrm{MoS}}_{2}$, ${\mathrm{MoSe}}_{2}$, ${\mathrm{WS}}_{2}$, and ${\mathrm{WSe}}_{2}$ monolayers, 
we find that the lowest-energy A exciton branches split, with the lower branch displaying parabolic behavior 
and the upper branch exhibiting non-parabolic behavior. We show that the EELS peaks unambiguously follow the non-parabolic upper band 
and thus has a linear dependence on $\mathbf{Q}$, indicating a direct connection of the dispersion relations of the EELS peaks
to the exciton band structure. We also find that EEL spectroscopy selectively probes only the bright exciton bands 
with their electric dipole being parallel to the direction of the transfered momentum.
This explains why only the upper band of the A exciton, which is a longitudinal exciton with
an in-plane dipole moment, was observed in the previous experiment \cite{JinhuaHong2020}.

This paper is organized as follows. In Sec. II, we briefly introduce the central theories of {\it ab initio} quasiparticle and exciton band structure calculations, 
followed by the computational details adopted in this work. In Sec. III, we carry out a thorough analysis of the widely used definitions of the EEL spectrum 
for quasi-2D systems, highlight the connections among these formulas, and provide a properly defined EEL spectrum formula for quasi-2D systems 
for the momentum and energy regimes studied in this work. In Sec. IV, we present the main results of the application of the theoretical methods outlined in 
Sec. II and Sec. III, covering the electronic and optical properties of monolayer ${\mathrm{WSe}}_{2}$, as discussed in Sec. IV A. In Sec. IV B, 
we discuss the exciton band structures in the small but finite $Q$ regime. In Sec. IV C, we present the calculated EEL spectra and examine 
the dispersion of the peaks in the EEL spectrum, and we discuss the exciton excitation mechanism by EELS. Finally, we summarize the results of this work in Sec. V.

\section{{\it Ab Initio} Bethe-Salpeter Equation Calculation}
All calculations were performed with the electronic structure code GPAW \cite{gpaw2024,gpaw2010,ase2017,JJ2005} 
using the accurate grid-based projector augmented-wave method (PAW) \cite{JJ2005}. 
\subsection{Structural relaxations}
For the structural relaxations we used density functional theory (DFT) within the generalized gradient approximation (GGA) with the Perdew-Burke-Ernzerhof (PBE) exchange-correlation functional \cite{Perdew1996}. For the optimizer we use the BFGS algorithm \cite{ase2017}. We model the ${\mathrm{WSe}}_{2}$ monolayer using the slab supercell method, employing a vacuum thickness of 15 $\textup{~\AA}$. We use a ${\rm 24 \times 24 \times 1}$ $\Gamma$-centered k-mesh sampling over the 2D Brillouin zone (BZ) and a plane-wave cutoff of 500 eV. 

\subsection{$\rm G_0W_0$ quasi-particle calculations}
The present $\rm G_0W_0$ quasi-particle (QP) calculations are performed by using the GPAW code \cite{gpaw2024,Huser2013}. 
The QP energies $E^{\mathrm{QP}}_{n\textbf{k}}$ are evaluated perturbatively by the linearized QP equation
\begin{equation}
E^{\mathrm{QP}}_{n\textbf{k}}=E^{\mathrm{KS}}_{n\textbf{k}}+Z_{n\textbf{k}}\langle{\psi^{\mathrm{KS}}_{n\bf{k}}|\Sigma_{\rm GW}+v_{x}-v_{xc}|\psi^{\mathrm{KS}}_{n\bf{k}}}\rangle
\label{Eq:GW}
\end{equation}
where $E^{\mathrm{KS}}_{n\bf{k}}$ and $\psi^{\mathrm{KS}}_{n\bf{k}}$ are the Kohn-Sham eigen-energies and wave functions. $Z_{n\textbf{k}}$ is the QP weight. $\Sigma$, $v_{x}$, and $v_{xc}$ are the $\rm GW$ self-energy, non-local exchange potential, and local exchange-correlation potential, respectively \cite{Huser2013,Rasmussen2016}. 

In the $\rm G_0W_0$ calculation, we adopt a $k$-mesh of ${\rm 24 \times 24 \times 1}$, 1200 bands, and a dielectric cutoff energy of 250 eV. 
We utilize 2D truncated Coulomb interactions \cite{Falco2013,Latini2015} between the ${\mathrm{WSe}}_{2}$ monolayer and its periodic replica. 
This approach is combined with the analytical treatment around the $\Gamma$-point to achieve faster $k$-points convergence when evaluating 
the screened interaction \cite{Rasmussen2016}. The $\rm G_0W_0$ self-energy is computed using the full frequency integration.

\subsection{The Bethe-Salpeter equation at finite momentum transfer}
The Bethe-Salpeter equation (BSE) describing the exciton with the center-of-mass momentum $\textbf{Q}$ reads \cite{Rohlfing2000,Yan2012}
\begin{equation}
\begin{aligned}
(E^{\mathrm{QP}}_{c\textbf{k+Q}}-E^{\mathrm{QP}}_{v\textbf{k}})A^{\lambda,\textbf{Q}}_{vc\textbf{k}}+\sum\limits_{v'c'\textbf{k}'}\langle{vc\textbf{k}\textbf{Q}|K^{eh}|v'c'\textbf{k}'\textbf{Q}}\rangle A^{\lambda,\textbf{Q}}_{v'c'\textbf{k}'}\\
=\Omega^{\lambda}(\textbf{Q})A^{\lambda,\textbf{Q}}_{vc\textbf{k}}   
\end{aligned}
\label{Eq:BSE}
\end{equation}
where $E^{\mathrm{QP}}_{c\textbf{k+Q}}-E^{\mathrm{QP}}_{v\textbf{k}}$ is the excitation energy of a free electron-hole pair, with the hole belonging to the band with index $v$ and the electron to the band with index $c$, forming a pair-orbital. $\Omega^{\lambda}(\textbf{Q})$ is the momentum-resolved exciton excitation energy of the $\lambda$-th state, and the corresponding exciton state can be expressed as the linear combination of the pair-orbitals with the expansion coefficients $A^{\lambda,\textbf{Q}}_{vc\textbf{k}}$, and the absolute square of the expansion coefficients, referred to as the exciton weight, represents the contributions of the pair-orbitals to the exciton wave function in reciprocal space. The electron-hole interaction is $ K^{eh}=K^{x}+K^{d}$. Here $K^{x}$ is the exchange bare Coulomb term \cite{Rohlfing2000,Yan2012}
\begin{equation}
\begin{aligned}
&\langle{vc\textbf{k}\textbf{Q}|K^{x}|v'c'\textbf{k}'\textbf{Q}}\rangle=\\
&\sum\limits_{\textbf{G}}\rho_{vc\textbf{k}}^{*}(\textbf{Q},\textbf{G})v_{\textbf{G}}(\textbf{Q})\rho_{v'c'\textbf{k}'}(\textbf{Q},\textbf{G}).
\end{aligned}
\label{Eq:e-h exchange}
\end{equation}
$K^{d}$ is the direct screened Coulomb term \cite{Rohlfing2000,Yan2012}
\begin{equation}
\begin{aligned}
&\langle{vc\textbf{k}\textbf{Q}|K^{d}|v'c'\textbf{k}'\textbf{Q}}\rangle=\\
&-\sum\limits_{\textbf{GG}'}\rho_{cc'\textbf{k}+\textbf{Q}}^{*}(\textbf{q},\textbf{G})W_{\textbf{GG}'}(\textbf{q})\rho_{vv'\textbf{k}}(\textbf{q},\textbf{G}')\  
\end{aligned}
\label{Eq:e-h direct}
\end{equation}
where $\rho_{vc\textbf{k}}(\textbf{Q},\textbf{G})=\langle{v\textbf{k}|e^{-i(\textbf{Q+G})\cdot\textbf{r}}|c\textbf{k+Q}}\rangle$ is the pair density, 
$\textbf{G}$ is the reciprocal lattice vector, and $\textbf{q}=\textbf{k}-\textbf{k}'$. $v_{\textbf{G}}(\textbf{Q})$ and $W_{\textbf{GG}'}(\textbf{q})$ 
are the bare Coulomb potential and the screened Coulomb potential in reciprocal space, respectively. 

In the Tamm-Dancoff approximation (TDA), we construct the density response function
in reciprocal space as \cite{Onida2002,Yan2012}
\begin{equation}
\begin{aligned}
& \chi_{\textbf{G}\textbf{G}'}(\textbf{Q},\omega)= \\
& \frac{1}{V_{c}}
\sum\limits_{\lambda}\sum\limits_{vc\textbf{k}}\sum\limits_{v'c'\textbf{k}'}\frac{A^{\lambda,\textbf{Q}}_{vc\textbf{k}}\rho_{vc\textbf{k}}(\textbf{Q},\textbf{G}) {\left [ {A}^{\lambda,\textbf{Q}}_{v'c'\textbf{k}'}\rho_{v'c'\textbf{k}'}(\textbf{Q},\textbf{G}') \right ]}^\ast}{\omega-\Omega^{\lambda}(\textbf{Q})+i\eta}
\end{aligned}
\label{Eq:Chi}
\end{equation}
where $V_{c}$ is the cell volume.

The exciton can be categorized into two distinct types—longitudinal and transverse excitons—depending 
on how its center-of-mass momentum aligns with its dipole matrix element \cite{Denisov1973}. 
The dipole matrix element of an exciton state $|{\lambda}\rangle$ is defined as 
\begin{equation}
\begin{aligned}
\textbf{P}_{\lambda}=\sum\limits_{vc\textbf{k}}A^{\lambda,\textbf{Q}}_{vc\textbf{k}}
\frac{\langle{v\textbf{k}|\textbf{p}|c\textbf{k+Q}}\rangle}{E^{\mathrm{QP}}_{c\textbf{k+Q}}-E^{\mathrm{QP}}_{v\textbf{k}}}
\end{aligned}
\label{Eq:dipole}
\end{equation}
where $\textbf{p}$ is the momentum operator. A longitudinal exciton refers to the case when the dipole matrix element aligns with the center-of-mass momentum $\textbf{Q}$, while for a transverse exciton they are perpendicular to each other \cite{Denisov1973}. At $\textbf{Q} \rightarrow 0$, Eq. (\ref{Eq:dipole}) represents the oscillator strength of an exciton state $|{\lambda}\rangle$ that constitutes the optical absorption spectrum \cite{Rohlfing2000}.

Our BSE calculations were performed on top of the DFT band structure and we shift the DFT band gap 
to match the ${\rm G_{0}W_{0}}$ band gap at the high-symmetric K point. We included 4 top valence 
bands and 4 lowest conduction bands to construct the BSE matrix, and the 2D truncated Coulomb 
interaction is used to avoid the spurious screening from the periodic images of the monolayer. 
The electronic screening was calculated using 100 bands with a dielectric cutoff of 50 eV. 
A dense $k$-mesh of $72\times72\times1$ is adapted for both the ground state and BSE calculations 
to achieve the high resolution in momentum transfer for the {\bf Q}-dependent EEL spectra. 

\section{THEORY of Electron energy loss spectra of quasi-2D materials}
\subsection{Optical absorbance}
\label{Optical absorbance}

The macroscopic dielectric function ($\varepsilon_{M}(\textbf{Q}, \omega)$) is the key quantity 
to obtain the optical properties of a 3D bulk system, which is evaluated as \cite{Onida2002}
\begin{equation}
\begin{aligned}
\varepsilon_{M}(\mathbf{Q}, \omega)= \frac{1}{\left.\varepsilon^{-1}_{\textbf{G}\textbf{G}'}(\textbf{Q},\omega)\right\vert_{\textbf{G}=0,\textbf{G}'=0}}
\end{aligned}
\label{Eq:matrix_eps}
\end{equation}
where $\varepsilon_{\textbf{G}\textbf{G}'}(\textbf{Q},\omega)$ is the dielectric matrix in reciprocal space, which related to $\chi_{\textbf{G}\textbf{G}'}(\textbf{Q},\omega)$ as 
\begin{equation}
\begin{aligned}
\varepsilon^{-1}_{\textbf{G}\textbf{G}'}(\textbf{Q},\omega) = 1 + v_{\textbf{G}}(\textbf{Q})\chi_{\textbf{G}\textbf{G}'}(\textbf{Q},\omega).
\end{aligned}
\label{Eq:eps_chi}
\end{equation}
Inverting the dielectric matrix is necessary to include the local field effects \cite{Onida2002}. However, it is cumbersome to evaluate all the dielectric matrix elements and perform the matrix inversion. Alternatively, we can obtain $\varepsilon_{M}(\textbf{Q}, \omega)$ as \cite{Onida2002}
\begin{equation}
\begin{aligned}
\varepsilon_{M}(\mathbf{Q}, \omega) = 1 - v_{0}(\textbf{Q})\bar\chi_{\mathbf{00}}(\textbf{Q},\omega)
\end{aligned}
\label{Eq:eps}
\end{equation}
where $\bar\chi_{\textbf{G}\textbf{G}'}(\textbf{Q},\omega)$ is the modified density response function in reciprocal space \cite{Onida2002,Yan2012}. Within the BSE formalism, $\bar\chi_{\textbf{G}\textbf{G}'}(\textbf{Q},\omega)$ has the same structure as $\chi_{\textbf{G}\textbf{G}'}(\textbf{Q},\omega)$ (see Eq. (\ref{Eq:Chi})), but the exciton eigenvectors and eigenvalues are obtained by using the modified BSE kernel \cite{Onida2002,Yan2012}, which adopts the modified bare Coulomb interaction $\bar v_{\textbf{G}}$ \cite{Onida2002}
\begin{equation}
\begin{aligned}
{\bar v_{\textbf{G}}(\textbf{Q})}=\left\{\begin{matrix}
0 & (${\rm if}$\ \textbf{G}=0)\\[6pt] 
v_{\textbf{G}}(\textbf{Q}) & (${\rm else}$) 
\end{matrix}\right.
\label{Eq:modified_Coulomb}
\end{aligned}
\end{equation}

For a quasi-2D system, a thick vacuum spacing between periodic quasi-2D systems has to be adopted in the supercell calculations 
to eliminate the spurious interlayer interaction. However, because $\bar\chi_{\mathbf{00}}$ is proportional to the inverse 
of $V_{c}$ \cite{Yan2012}, with the increasing vacuum thickness, $\varepsilon_{M}(\textbf{Q}, \omega)$ in Eq. (\ref{Eq:eps}) 
will tend to unity, indicating the physical properties of a quasi-2D system will be overwhelmed by that of the vacuum layer. 
To be able to reduce the size of the supercell in the $z$-direction, a truncated Coulomb interaction can be used \cite{Rubio2006}. 
In 3D bulk materials, the bare Coulomb interaction $v_{\textbf{G}}(\textbf{Q}) = 4\pi/|\textbf{Q}+\textbf{G}|^{2}$. 
In quasi-2D materials, we replace $v_{\textbf{G}}(\textbf{Q})$ with the truncated Coulomb interaction $v^{trunc}_{\textbf{G}}(\textbf{Q})$, 
which is given by \cite{Rubio2006,Ismail2006} 
\begin{equation}
\begin{aligned}
v^{trunc}_{\textbf{G}}(\textbf{Q}) = \frac{4\pi}{|\textbf{Q}+\textbf{G}|^{2}}[1-e^{-|\textbf{Q}_{||}+\textbf{G}_{||}|d/2} \cos(G_{z}d/2)]
\label{Eq:trunc_Coulomb}
\end{aligned}
\end{equation}
where $d$ is the length of the supercell in the non-periodic direction, and the truncation length is set to $d/2$. However, 
since $v^{trunc}_{0}(\mathbf{Q}\rightarrow 0) \approx 2\pi d/|\textbf{Q}|$, a straightforward use of Eq. (\ref{Eq:eps}) 
would also result in a vanishing absorption spectra. 
Therefore, we define the polarizability as \cite{Cudazzo2011,gpaw2024}
\begin{align}
{\rm \alpha}(\omega)=-\lim_{|\textbf{Q}| \rightarrow 0} \frac{d}{|\textbf{Q}|^{2}}\bar\chi_{\mathbf{00}}(\textbf{Q},\omega).
\label{Eq:polarizability}
\end{align}
If a truncated Coulomb interaction is used to evaluate $\bar\chi_{\mathbf{00}}(\textbf{Q},\omega)$, this polarizability remains finite 
even a thick vacuum is adopted. Equation (\ref{Eq:polarizability}) measures the induced dipole momentum per area of 
a quasi-2D system \cite{Cudazzo2011,gpaw2024}. The imaginary part of the polarizability represents the optical absorbance \cite{Cudazzo2011,gpaw2024}.

\subsection{Electron energy-loss spectrum} \label{Electron energy-loss spectrum}

\subsubsection{Different EEL spectrum formulas}
The momentum-dependent EEL spectrum is obtained experimentally through transmission electron energy-loss spectroscopy \cite{Abajo2010,JinhuaHong2020}. The quantity that is directly measured in the electron energy-loss spectroscopy experiments is the doubly differential cross-section $(\partial^2 \sigma / \partial \Omega \partial \omega)$ \cite{Sturm1993}, which represents the relative probability for the incident electron experiencing a scattering into the solid angle $(d\Omega)$ and energy interval $(\omega \rightarrow \omega+d\omega)$. The doubly differential cross section is \cite{FINK1989121} 
\begin{equation}
\begin{aligned}
\frac{\partial^2 \sigma}{\partial \Omega \partial \omega} = (\frac{\partial \sigma}{\partial \Omega})_{Ruth}S(\textbf{Q},\omega)
\end{aligned}
\label{Eq:DDSC}
\end{equation}
where $\mathbf{Q}$ is the total momentum transfer of an incident electron. The doubly differential cross-section is the product of two terms. The first term $(\partial \sigma / \partial \Omega)_{Ruth} = 4/(a_{B}|\textbf{Q}|^4)$ is the elastic Rutherford scattering cross-section, where $a_B$ is the Bohr radius. This term describes the elastic differential cross-section of two charged particles, with the interaction between them being the Coulomb force. The other term $S(\textbf{Q},\omega)$ is the dynamical structure factor, containing the information on the response of the electrons in the system to the external perturbation \cite{FINK1989121}. $S(\textbf{Q},\omega)$ in 3D is related to the inverse dielectric function in reciprocal space as \cite{Sturm1982} 
\begin{equation}
\begin{aligned}
S(\textbf{Q},\omega) \equiv -\frac{|\textbf{Q}|^2}{4\pi^2e^2}{\rm Im}[\varepsilon^{-1}_{00}(\textbf{Q},\omega)].
\end{aligned}
\label{Eq:structure_factor}
\end{equation}
Since $S(\textbf{Q},\omega) \sim |\textbf{Q}|^2$ for small $\textbf{Q}$, we define the EEL spectrum ($L(\mathbf{Q}, \omega)$) as the dimensionless quantity
\begin{equation}
\begin{aligned}
L(\mathbf{Q}, \omega) = \frac{4\pi^2e^2}{|\textbf{Q}|^2}S(\textbf{Q},\omega) \equiv -{\rm Im}[\varepsilon^{-1}_{00}(\textbf{Q},\omega)].
\end{aligned}
\label{Eq:eels_def}
\end{equation}
However, from the discussion of the absorption spectra in the previous section, it is evident that a straight-forward calculation 
of EEL spectrum from the dielectric function will give a vanishing EEL spectrum as $\mathbf{Q}\rightarrow 0$ in quasi-2D materials. 
There are different strategies to solve this problem in the literature, which yield different definitions of the EEL spectrum. The two most commonly used definitions are listed as $Definition \:1$ and $Definition \:2$ below.
\\
\\
$Definition \:1 :$ The EEL spectrum is conventionally obtained from taking the imaginary part of the inverse of the macroscopic dielectric function ($\varepsilon_{M}(\textbf{Q}, \omega)$) analogously to the 3D case \cite{Sturm1982,Sturm1993}
\begin{equation}
\begin{aligned}
&L(\textbf{Q},\omega) = {\rm -Im[1/\varepsilon_{M}}(\textbf{Q},\omega)] \\
&= \frac{{\rm Im[\varepsilon_{M}(\textbf{Q}, \omega)]}}{({\rm Re[\varepsilon_{M}(\textbf{Q}, \omega)]})^2+({\rm Im[\varepsilon_{M}(\textbf{Q}, \omega)]})^2}.
\end{aligned}
\label{Eq:eels_tradition}
\end{equation}
This definition of EEL spectrum is equivalent to the 3D formula in Eq. (\ref{Eq:eels_def}), and it has been widely used also for quasi-2D materials in the literature \cite{Liou2015,JinhuaHong2021,Priya2011,XWZhao2020,Guang2023}. However, as discussed above $\varepsilon_{M} \rightarrow 1$ for large supercells, or even for finite supecells if a truncated Coulomb interaction is used to evaluate the dielectric matrix. 
Therefore, this formula can only be used without Coulomb truncation and for finite supercells and is in principle not applicable to quasi-2D systems. To get a reasonable definition in quasi-2D systems, one can evaluate $\varepsilon_M$ from Eq. (\ref{Eq:eps}) where $\bar{\chi}_{\mathbf{00}}$ is evaluated with a truncated Coulomb interaction. However, the matrix inversion formula used to derive Eq. (\ref{Eq:eps}) relies on the assumption that the Coulomb interaction used to calculate $\chi_{\textbf{G}\textbf{G}'}(\mathbf{Q})$ in Eq. (\ref{Eq:eps_chi}) is the same as the $v_{0}(\mathbf{Q})$ in the product ($v_{0}(\mathbf{Q})\bar\chi_{\mathbf{00}}(\textbf{Q},\omega)$). Using a \emph{untruncated interaction} $v_{0}(\mathbf{Q})=4\pi/|\mathbf{Q}|^2$ but a \emph{truncated interaction} $v^{trunc}_{\mathbf{G}}(\mathbf{Q})$ to evaluate $\chi_{\textbf{G}\textbf{G}'}(\mathbf{Q})$ is thus an approximation which, as we will see, has a relatively large impact on the results. 
When we refer to \emph{Definition 1} below, we will always consider the case where $\varepsilon_M$ is evaluated from Eq. (\ref{Eq:eps}) and $\bar{\chi}$ is evaluated with a truncated Coulomb interaction.
To avoid the scaling of the spectra with the supercell size, Liou $et\,al.$ \cite{Liou2015} introduced the effective volume approach by replacing the supercell volume with half of the bulk counterpart to calculate the EEL spectra of graphene.
\\
\\
$Definition \:2 :$ The EEL spectrum has also been defined to be related to the imaginary part of the density response function $\chi_{\textbf{0}\textbf{0}}(\textbf{Q},\omega)$ as \cite{Nerl2017}
\begin{equation}
\begin{aligned}
&L(\textbf{Q},\omega)=-\frac{4\pi}{|\textbf{Q}|^{2}}\text{Im}\chi_{\mathbf{00}}(\textbf{Q},\omega).
\end{aligned}
\label{Eq:eels_chi00}
\end{equation}
For 3D bulk systems, this definition is equivalent to Eqs. (\ref{Eq:eels_def})-(\ref{Eq:eels_tradition}). For quasi-2D systems, 
despite $\chi_{\textbf{0}\textbf{0}}(\textbf{Q},\omega) \propto$ $1/V_{c}$, we note that Eq. (\ref{Eq:eels_chi00}) does not suffer 
from the supercell size problem or the need to use an approximate matrix inversion formula (like \emph{Definition 1} above) 
since $\chi$ can be directly evaluated with a truncated Coulomb interaction. On the other hand, to get a quantity 
which is independent of the supercell size, Eq. (\ref{Eq:eels_chi00}) can be multiplied by $d$, which is the size 
of the supercell in the $z$-direction (similar to the polarizability in the previous section) 
\begin{equation}
\begin{aligned}
&\tilde{L}(\textbf{Q},\omega)=-\frac{4\pi d}{|\textbf{Q}|^{2}}\text{Im}\chi_{\mathbf{00}}(\textbf{Q},\omega).
\end{aligned}
\label{Eq:eels_chi00_proper}
\end{equation}
As will be shown in the following section, Eq. (\ref{Eq:eels_chi00_proper}) can be formally derived as a limiting case of the integrated EELS formula proposed in Ref. \cite{nazarov2015electronic}.
However, since $d$ is just an overall multiplicative factor, and the experimental spectra is typically normalized, the additional factor $d$ is typically not important in practical calculations.
Eq. \ref{Eq:eels_chi00} is implemented in the GPAW code \cite{gpaw2024}, and has been adopted in the previous works on quasi-2D systems \cite{Yan2011,Andersen2013,Nerl2017,KST2017}.
\\

Both definitions above rely on an analogy to the 3D case and only depends on the macroscopic components ($\mathbf{G}=\mathbf{G}'=0$) 
of the response functions. However, for quasi-2D systems which are confined in the $z$-direction it is not immediately clear 
that the finite $G_z$ components should vanish. Actually, as shown in Ref. \cite{nazarov2015electronic} and discussed 
under \emph{Definition 3} below, all $G_z$ components of $\chi$ contribute to the EEL spectrum.
\\
\\
$Definition \:3 :$ Starting from Eq. (\ref{Eq:eels_def}) with an explicit dynamical structure factor $S$ applicable to a quasi-2D system, Nazarov \cite{nazarov2015electronic} showed that the relevant quantity for the doubly differential cross-section in EELS for quasi-2D materials is what we here will call the integrated EELS
\begin{equation}
\begin{aligned}
\tilde{L}(\mathbf{Q}, \omega) = -\frac{4\pi}{|\textbf{Q}|^{2}}\mathrm{Im} \int_{-\infty}^{\infty} e^{iq_z(z'-z)}\chi_{\mathbf{00}}(z, z', \mathbf{Q}_{||},\omega)dzdz'
\end{aligned}
\label{Eq:Q2DEELS}
\end{equation}
where $\mathbf{Q}$ is the total momentum transfer of a incident electron consisting of its in-plane component $\mathbf{Q}_{||}$ and out-of-plane component $q_z\mathbf{\widehat{z}}$. $\chi_{\mathbf{G}_{||}\mathbf{G}'_{||}}(z, z', \mathbf{Q}_{||},\omega)$ is the in-plane Fourier transform of the density response function \cite{Shiwu2009}. This formula is rigorous but difficult to implement.

In the following section, we will demonstrate that Eq. (\ref{Eq:Q2DEELS}) proposed by Nazarov \cite{nazarov2015electronic} will actually reduce to Eq. (\ref{Eq:eels_chi00_proper}) within the energy and momentum-transfer regimes discussed in this article.

\subsubsection{Justification for the EELS formula used in this work}
To construct the formula of EEL spectrum for quasi-2D systems without relying on $\varepsilon_{M}$, tailored to fit the energy and momentum-transfer regimes discussed in this article, we start from the EEL spectrum formula proposed by Nazarov \cite{nazarov2015electronic} (\emph{Definition 3})
\begin{equation}
\begin{aligned}
\tilde{L}(\mathbf{Q}, \omega) = -\frac{4\pi}{|\textbf{Q}|^{2}}\mathrm{Im} \int_{-\infty}^{\infty} e^{iq_z(z'-z)}\chi_{\mathbf{00}}(z, z', \mathbf{Q}_{||},\omega)dzdz'.
\end{aligned}
\label{Eq:Q2DEELS_2}
\end{equation}

Firstly, $\chi_{\mathbf{G}_{||}\mathbf{G}'_{||}}(z, z', \mathbf{Q}_{||},\omega)$ is related to the density response function in reciprocal space $\chi_{\mathbf{GG}'}(\mathbf{Q}_{||},\omega)$ as \cite{Shiwu2009}
\begin{equation}
\begin{aligned}
\chi_{\mathbf{\mathbf{G}_{||}\mathbf{G}'_{||}}}(z, z', \mathbf{Q}_{||},\omega)=\frac{1}{d}\sum\limits_{ G_{z}G'_{z}}e^{iG_{z}z}\chi_{\mathbf{GG}'}(\mathbf{Q}_{||},\omega)e^{-iG'_{z}z'}
\end{aligned}
\label{Eq:Chi_Fourier}
\end{equation}
where $d$ is the length of the unit cell in the out-of-plane direction $z$. Next, the expression for 
$q_z$ is \cite{nazarov2015electronic} 
\begin{align}
q_z = \frac{\omega + \frac{\hbar}{2m}{\left | \mathbf{Q}_{||} \right | }^2}{v} 
\label{Eq:qz}
\end{align}
where $v$ and $m$ are the velocity and mass of the incident electron, respectively. 
In practice, $|\mathbf{Q}_{||}| \gg q_z$ due to the high velocity of the incident electrons \cite{Abajo2010,JinhuaHong2020}. 
By exploiting the fact that $\chi_{\mathbf{G}_{||}\mathbf{G}'_{||}}(z, z', \mathbf{Q}_{||},\omega)$ would practically only be nonzero for $-d/2 < z < d/2$ and $-d/2 < z' < d/2$, we can relate $\tilde{L}(\mathbf{q}, \omega)$ and $\chi_{\mathbf{GG}'}(\mathbf{Q}_{||},\omega)$ via Eq. (\ref{Eq:Q2DEELS}) and Eq. (\ref{Eq:Chi_Fourier}) as
\begin{equation}
\begin{aligned}
& \tilde{L}(\mathbf{Q}, \omega) \\
& = -\frac{4\pi}{|\textbf{Q}|^{2}d}\sum\limits_{ G_{z}G'_{z}} \mathrm{Im}   
\chi_{\mathbf{0}G_z\mathbf{0}G'_z}(\mathbf{Q}_{||},\omega)\frac{2-2cos(q_zd)}{(G_z-q_z)(G'_z-q_z)}.
\end{aligned}
\label{Eq:EELS_decompose}
\end{equation}

Note that $q_zd$ is practically a small number because of the high velocity of the normally incident electrons \cite{Abajo2010,JinhuaHong2020}. Consequently, performing the Taylor expansion of Eq. (\ref{Eq:EELS_decompose}) up to the lowest order in $\mathbf{Q}$ yields

\begin{equation}
\begin{aligned}
&\tilde{L}(\mathbf{Q}, \omega) = -\frac{4\pi d}{|\textbf{Q}|^{2}}\mathrm{Im}\chi_{\mathbf{0}\mathbf{0}}(\mathbf{Q}_{||},\omega) \\
&-\frac{4\pi d}{|\textbf{Q}|^{2}}\sum\limits_{G_{z},G'_{z}\neq 0} \mathrm{Im} 
\chi_{\mathbf{0}G_z\mathbf{0}G'_z}(\mathbf{Q}_{||},\omega)\frac{q_z^2}{(G_z-q_z)(G'_z-q_z)}.
\end{aligned}
\label{Eq:EELS_taylor}
\end{equation}

The $G_z=G'_z=0$ component dominates the EEL spectrum, while the contributions of other $G_z\neq 0$ or $G'_z\neq 0$ terms in Eq. (\ref{Eq:EELS_taylor}) are significantly smaller because $q_z \ll G_z$ and $q_z \ll  G'_z$ in the energy and momentum transfer regions that we focus on in this work. 
As a result, within the limit of $q_zd \ll 1$ and noting that $|\mathbf{Q}_{||}| \gg q_z$, Eq. (\ref{Eq:Q2DEELS}) reduces to

\begin{equation}
\begin{aligned}
&\tilde{L}(\textbf{Q}_{||},\omega)=-\frac{4\pi d}{|\textbf{Q}_{||}|^{2}}\text{Im}\chi_{\mathbf{00}}(\textbf{Q}_{||},\omega).
\end{aligned}
\label{Eq:eels_chi00_simplified}
\end{equation}

This formula is exactly the same as Eq. (\ref{Eq:eels_chi00_proper}). Eqs. (\ref{Eq:eels_chi00}) and (\ref{Eq:eels_chi00_proper}) 
has already been adopted in several previous papers on the quasi-2D systems \cite{Yan2011,Andersen2013,Nerl2017,KST2017}. 
However, rather than motivating it from an analogy with the 3D case, we derive it from the rigorous formalism of EEL spectrum 
(i.e., Eq. (\ref{Eq:Q2DEELS})). Our derivation shows that \emph{Definition 2} is valid in the low-$\mathbf{Q}$ region with 
the standard electron energies used in EELS. For the electron energies in Ref. \cite{JinhuaHong2020}, the upper bound of $\mathbf{Q}_{||}$ 
is $\mathbf{Q}_{||} \ll 8.8 \textup{~\AA}^{-1}$ (see supplementary note 1 in the Supplemental Material (SM) \cite{Stone-SM}).
For lower electron energies and higher $\mathbf{Q}$ the full formula (\emph{Definition 3}) has to be used.  
Furthermore, the other commonly used formula \emph{Definition 1} cannot be derived in any rigorous manner and should therefore 
be considered as an {\it ad hoc} procedure. As we will show below this {\it ad hoc} formula also yields worse agreement with experiment. 

It is worth noting that for a quasi-2D system, $\bar\chi_{\mathbf{00}}(\textbf{Q},\omega)$ is equivalent 
to $\chi_{\mathbf{00}}(\textbf{Q},\omega)$ as $\textbf{Q} \rightarrow 0$ \cite{Nerl2017,JinhuaHong2022}. 
Therefore, the optical absorbance has the same structure as the EEL spectrum \cite{Nerl2017,JinhuaHong2022}. 
This is a unique characteristic of a quasi-2D system, as explained below. The $\textbf{G}=0$ component 
of the electron-hole exchange interaction $K^{x}$ (Eq. \ref{Eq:e-h exchange}) consists of the pair 
density $\rho_{vc\textbf{k}}(\textbf{Q},\textbf{0})$ and the Coulomb potential $v_{\textbf{0}}(\textbf{Q})$.
As $\textbf{Q} \rightarrow 0$, $\rho_{vc\textbf{k}}(\textbf{Q},\textbf{0})$ is proportional 
to $|\textbf{Q}|$ \cite{Yan2011,Yan2012} for both 3D bulk and quasi-2D systems. For quasi-2D systems, the Coulomb potential 
behaves as $v_{\textbf{0}}(\textbf{Q}) \sim  1/|\textbf{Q}|$ after applying Coulomb truncation \cite{Rubio2006,Ismail2006}. 
This implies that the $\textbf{G}=0$ term of $K^{x}$ scales as $|\textbf{Q}|$, making it negligible in 
the optical limit ($\textbf{Q} \rightarrow 0$). Consequently, the full BSE kernel is reduced to the modified BSE 
kernel \cite{Nerl2017,JinhuaHong2022}, indicating that $\bar\chi_{\mathbf{00}}$ is equivalent to $\chi_{\mathbf{00}}$. 
This behavior is different for 3D bulk systems, where the Coulomb potential scales as $1/{|\textbf{Q}|^2}$.

\section{Application to WSe$_2$ monolayer}
We begin by presenting the GGA and $\rm G_{0}W_{0}$ quasi-particle band structure of the ${\mathrm{WSe}}_{2}$ monolayer 
to determine the effective masses and scissor shift. We then show the optical absorbance obtained within the 
scissor-corrected BSE (SC-BSE) framework, identifying five exciton states. Subsequently, we analyze the exciton band structures 
of the A exciton and its dark counterpart in the small {\bf Q} regime. Following this, we calculate the EEL spectra 
and present the dispersion curves of the EELS peaks. Finally, we discuss the exciton excitation mechanism 
in EELS to explain these dispersion curves.

\subsection{Electronic and optical properties}

\begin{table*}[t]
\caption{The electronic band gaps at the K-K transition ($E_{KK}$) and $\rm K - \Lambda$ transition ($E_{K\Lambda}$). $\Delta_{K-\Lambda}$ 
is the energy difference between $E_{KK}$ and $E_{K\Lambda}$ ($\Delta_{K-\Lambda} = E_{K\Lambda} - E_{KK}$).
The SOC-induced energy splitting between the first two topmost valence bands ($\Delta_{SOC}^{v}$) and the two lowest conduction bands 
($\Delta_{SOC}^{c}$) at the K point. The effective masses (in unit of free electron mass $m_{0}$) of the highest valence band ($m_{v}$), 
lowest and the second lowest conduction bands ($m_{c_{1}}$ and $m_{c_{2}}$). The available experimental data (Expt.) are listed for comparison.}
\begin{ruledtabular}
\begin{tabular}{l c c c c c c c c}
 & \multicolumn{1}{c}{$E_{KK}$ {\rm(eV)}} & \multicolumn{1}{c}{$E_{K\Lambda}$ {\rm(eV)}}& \multicolumn{1}{c}{$\Delta_{K-\Lambda}$ {\rm(meV)}}& \multicolumn{1}{c}{$\Delta_{SOC}^{v}$ {\rm(meV)}} & \multicolumn{1}{c}{$\Delta_{SOC}^{c}$ {\rm(meV)}} & \multicolumn{1}{c}{$m_{v}$ ($m_{0}$)} & \multicolumn{1}{c}{$m_{c_{1}}$ ($m_{0}$)} & \multicolumn{1}{c}{$m_{c_{2}}$ ($m_{0}$)} \\
 \hline
 GGA    & 1.236 & 1.283 & 47 & 490 & 42.1 & 0.330 & 0.391 & 0.264 \\
 $\rm G_{0}W_{0}$     & 2.067 & 2.007 & -60 & 488 & 39.5 & 0.333 & 0.390 & 0.264 \\
 Expt.   & 2.2\footnote{Reference \cite{Chendong2015}}, 1.95\footnote{Reference \cite{Yi2016}}  & 2.12\footnotemark[1] & -80\footnotemark[1] & 475\footnotemark[2], 513\footnote{Reference \cite{Le2015}} & 14\footnote{Reference \cite{Piotr2021}}, $12\pm0.5$\footnote{Reference \cite{Ren2023}} & 0.35\footnotemark[3] & ... & ...\\
\end{tabular}
\end{ruledtabular}
\label{table:QPband}
\end{table*}

\begin{figure}[tbph] 
    \centering
    \includegraphics[width=8.5cm]{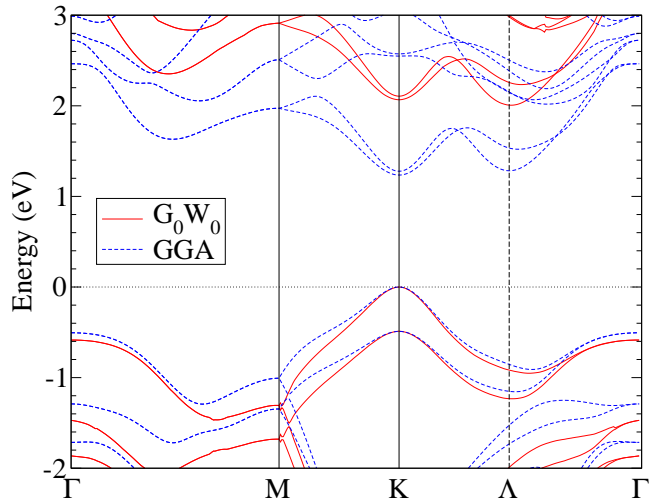}
    \caption{Quasi-particle band structure of the ${\mathrm{WSe}}_{2}$ monolayer from our $\rm G_{0}W_{0}$ (solid red lines) and GGA (blue dashed lines) calculations. The valence band maximum, located at the K point, is shifted to 0 eV.}
    \label{Fig:band}
\end{figure}

Figure \ref{Fig:band} shows the GGA and $\rm G_{0}W_{0}$ quasi-particle band structures of the ${\mathrm{WSe}}_{2}$ monolayer. 
Our GGA calculation indicates that the ${\mathrm{WSe}}_{2}$ monolayer is a semiconductor with the $\rm K - K$ direct band gap 
($E_{KK}$) of 1.236 eV, which underestimates the experimental values (see Table \ref{table:QPband}). In contrast, the $\rm G_{0}W_{0}$ 
quasi-particle band gap exhibits a significant improvement and aligns well with the experimental values, indicating the importance 
of $\rm G_{0}W_{0}$ quasi-particle calculations. In accordance with previous calculations \cite{Matthias2018,Ramasubramaniam2012} 
as well as experimental measurements \cite{Chendong2015}, the $\rm G_{0}W_{0}$ bandstructure reveals that the ${\mathrm{WSe}}_{2}$ 
monolayer has a $\rm K - \Lambda$ indirect quasi-particle band gap ($E_{K\Lambda}$) of 2.007 eV with the conduction band minimum 
at the $\rm \Lambda$ point lying midway between the $\rm \Gamma$ and K points. The difference between the $\rm K - K$ band gap 
and $\rm K - \Lambda$ band gap, defined as $\Delta_{K-\Lambda} = E_{K\Lambda} - E_{KK}$, is -60 meV, agrees well with 
the experimental measurements \cite{Chendong2015,Hsu2017} (see Table \ref{table:QPband}).
Additionally, the significant spin-orbit coupling (SOC) effect in the ${\mathrm{WSe}}_{2}$ monolayer, attributed to 
the heavy tungsten element, leads to the substantial valence and conduction bands splitting of 488 meV and 39.5 meV at the K valleys. 
The SOC-induced valence band splitting is consistent with the experimental findings (see Table \ref{table:QPband}), 
and the relatively small conduction band splittings are slightly larger than the experimental measurements. We also list 
the effective masses (in unit of free electron mass $m_{0}$) of the highest valence band ($m_{v}$), lowest and 
second lowest conduction bands (labeled as $m_{c_{1}}$ and $m_{c_{2}}$). The effective masses are obtained 
by fitting the electronic bands centered at the K point with a third-order polynomial. The fitting range spans 
$\textbf{Q} = 0.05  \textup{~\AA}^{-1}$ (within 3 $\%$ of the size of Brillouin zone) in the $\Gamma$M direction. 
Notably, the differences between the effective masses obtained from the GGA and $\rm G_{0}W_{0}$ band structures are small 
(within 1 $\%$ differences).  In the following scissor-corrected Bethe-Salpeter equation (SC-BSE) calculations, 
we shift the GGA $\rm K - K$ direct band gap to align it with that of the $\rm G_{0}W_{0}$ QP band structure. 
We find that the difference between the GGA+scissor and the $\rm G_{0}W_{0}$ band structures in the vicinity of 
the K point is small, as shown in Fig. S5 of the SM~\cite{Stone-SM}, indicating the scissor correction is a good approximation 
for describing the exciton states centered around the K valleys.

\begin{figure}[tbph]
    \centering
    \includegraphics[width=9cm]{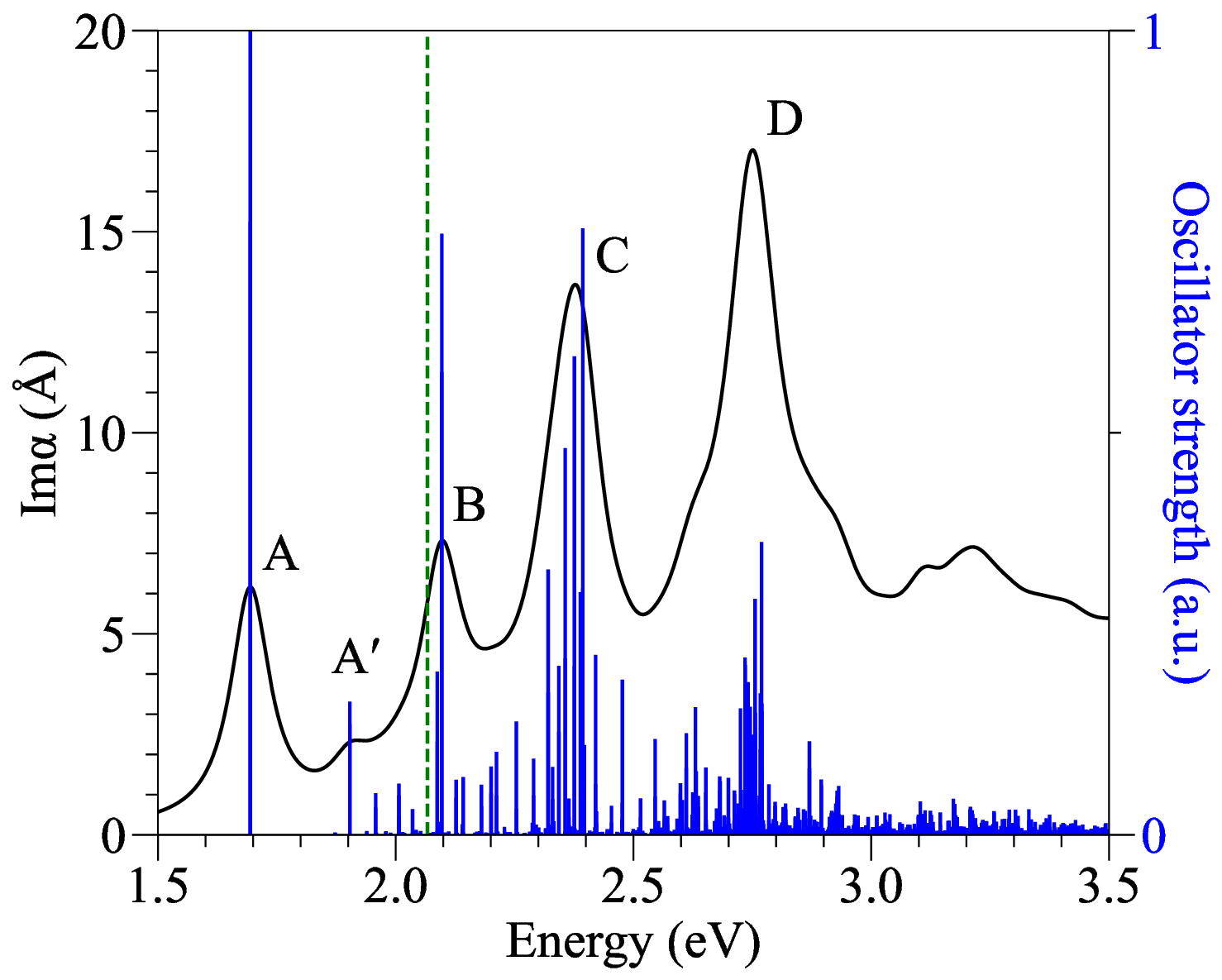}
    \caption{The imaginary part of the polarizability of the ${\mathrm{WSe}}_{2}$ monolayer calculated within the SC-BSE 
framework (solid black line) and the normalized oscillator strengths (blue vertical lines) of each exciton. 
The first four pronounced bright exciton peaks are labeled by A, B, C, and D. The peak with small intensity between the A and B peaks 
is labeled by $\rm A'$. The dashed green line indicates the quasi-particle band gap.}
    \label{Fig:pol}
\end{figure}
Figure \ref{Fig:pol} illustrates the imaginary part of polarizability of the ${\mathrm{WSe}}_{2}$ monolayer calculated within the SC-BSE framework using Eq. (\ref{Eq:polarizability}) (black solid line) and the normalized oscillator strengths (blue vertical lines) of each exciton. The quasi-particle band gap is also shown (green dashed line). The strong excitonic effects of the ${\mathrm{WSe}}_{2}$ monolayer significantly alter the absorption spectrum in the low-energy window. In the quasi-particle band gap, a prominent peak (labeled A) corresponds to the lowest-energy bright exciton state with the highest oscillator strength. Figure \ref{Fig:BZ}(b) illustrates the dominant weight of the A exciton on top of the spin-resolved DFT+scissor band structure. The A exciton originates from the pair-orbitals comprised of the topmost valence band and the second lowest conduction band having the same spin orientation (like-spin transition states) and are located at the K and $\rm K'$ valleys.
The reduced electronic screening effect in the ${\mathrm{WSe}}_{2}$ monolayer results in the large binding energy of 0.374 eV which agrees well with the measured values (see TABLE \ref{table:exciton exp}). The significant SOC-induced energy splitting between the first two highest valence bands introduces another bright exciton state labeled B, which contributes to the intense peak located slightly above the QP band gap. The dominant weight of the B exciton is primarily located at the like-spin transition between the second highest valence band and the lowest conduction band (see Fig. S6(a)). The large exciton excitation energy separation between the A and B excitons leads to the observation of the less intense peak (labeled $\rm A'$) between A and B peaks, corresponding to an exciton state in the QP band gap with the binding energies of 0.164 eV. The dominant weight of the $\rm A'$ exciton is primarily located at the same like-spin transition as the A exciton (see Fig. S6(b)). This indicates $\rm A'$ exciton is one of the excited states of the A exciton Rydberg series. Unlike the A, $\rm A'$, and B peaks, the high-energy C and D peaks consist of numerous exciton states with finite oscillator strength, resulting in the high spectrum intensity.

\subsection{Exciton band structure}
\begin{table*}[t]
\caption{Exciton excitation energy ($\Omega(0)$) and exciton binding energy ($E_b$) for various bright exciton states of ${\mathrm{WSe}}_{2}$ monolayer. The available experimental data (Expt.) are also listed for comparison.}
\begin{ruledtabular}
\begin{tabular}{l c c c c c}
 & $\Omega(0)$ (eV) & $E_b$ (eV)& \multicolumn{3}{c}{$\Omega(0)$ (eV)} \\
 \cline{2-3}\cline{4-6}
 & \multicolumn{2}{c}{Exciton A ($\rm A'$)} & Exciton B & Exciton C & Exciton D \\
 \hline
 SC-BSE & 1.693 (1.903)  &  0.374 (0.164) & 2.097  & 2.374  & 2.748 \\
 Expt.   & 1.65\footnote{Reference \cite{Mak2014}}, 1.67\footnote{Reference \cite{Tony2018}}, 1.661\footnote{Reference \cite{Hsu2017}}  & 0.37\footnotemark[1], 0.315\footnotemark[2] & 2.08\footnotemark[1]$^{,}$\footnotemark[2] & 2.43\footnote{Reference \cite{Tony2014}} & 2.89\footnotemark[4] \\
\end{tabular}
\end{ruledtabular}
\label{table:exciton exp}
\end{table*}

\begin{figure*}[tbph]
    \centering
    \includegraphics[width=17.5cm]{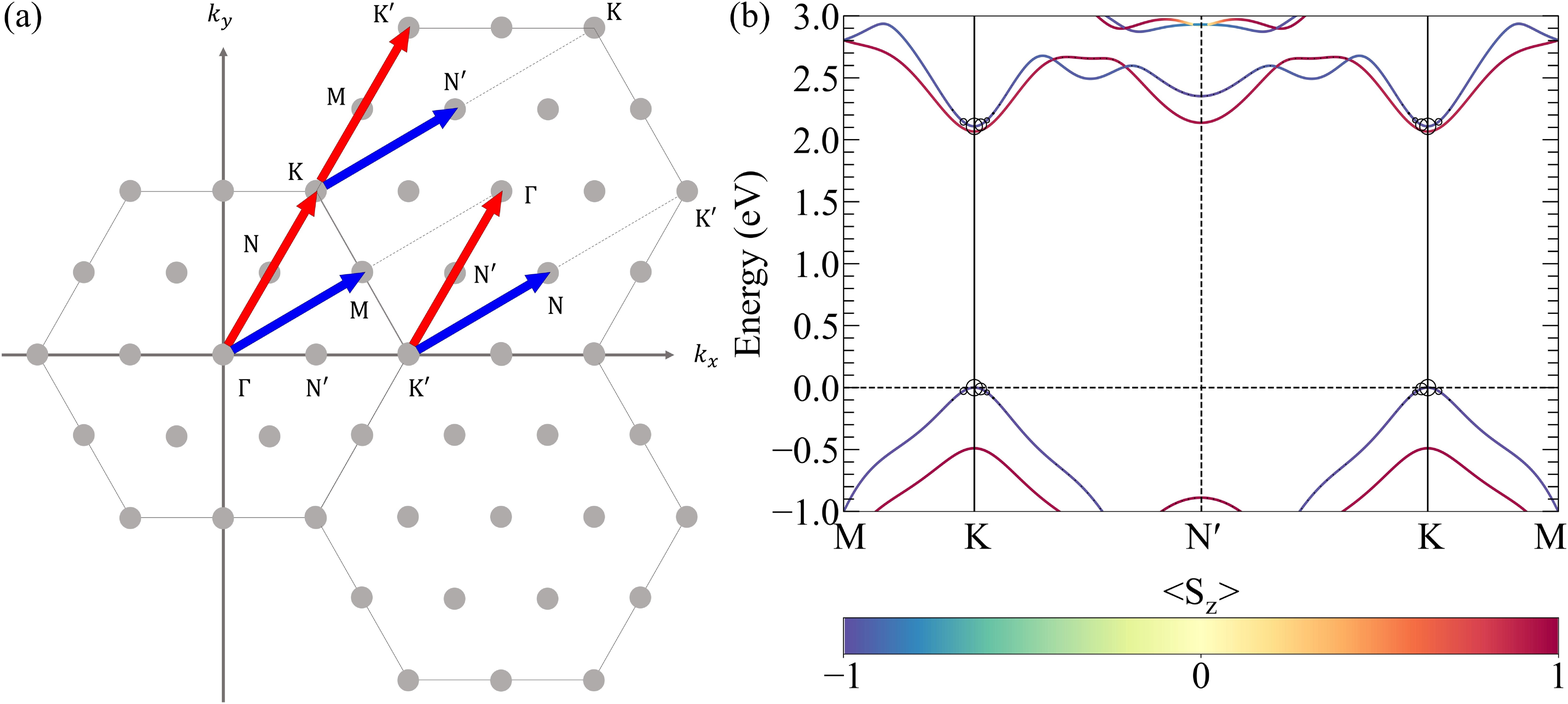}
    \caption{(a) The two-dimensional hexagonal Brillouin zone. The special points are labeled, and they are denoted by gray dots. The blue and red arrows indicate the momentum transfers along $\Gamma$M and $\Gamma$K directions. These arrows are then shifted to the K and $\rm K'$ points, indicating the corresponding transitions for excitons that are localized at the K and $\rm K'$ points. (b) The spin-resolved DFT+scissor band structure spans from one K point to another, passing through the middle N point. Colors from blue to red represent the expectation value of spin projection in z direction, indicating the spin polarization. We also show the A exciton weight (black circles) of the pair-orbitals that have the dominant contributions.}
    \label{Fig:BZ}
\end{figure*}

\begin{figure}[tbph]
    \centering
    \includegraphics[width=8.5cm]{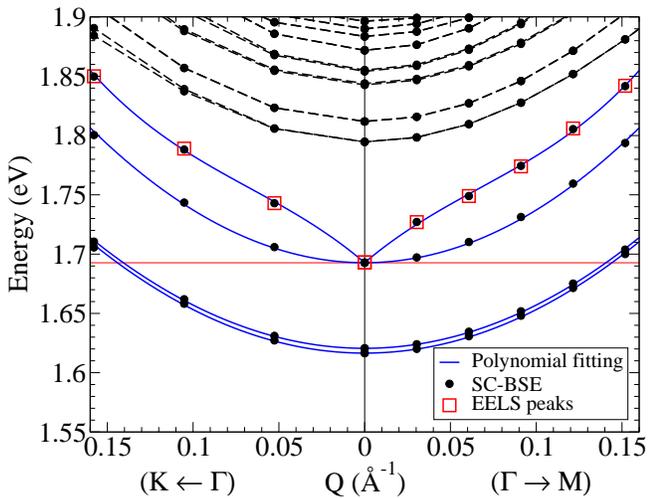}
    \caption{The exciton band structure and the calculated EELS peaks dispersion of the A exciton of the ${\mathrm{WSe}}_{2}$ monolayer. The black dots are the exciton excitation energies by solving SC-BSE at different $\textbf{Q}$, and the red squares are the lowest A EELS peak positions. The four lowest exciton bands are represented by a polynomial fit in $\Gamma$M direction (blue lines), while the rest exciton excitation energies are connected by black dashed lines serving as a guide to the eye. The excitation energy of the A exciton states at $\textbf{Q}=0$ is indicated by the horizontal red line.}
    \label{Fig:exciton_band}
\end{figure}

The excitation energy of an exciton $\Omega^{\lambda}$ depends on its center-of-mass momentum $\textbf{Q}$, 
and the relation between $\Omega^{\lambda}$ and $\textbf{Q}$ is referred to as the exciton band structure, 
which contains essential information of the exciton dynamics. 
Figure \ref{Fig:BZ}(a) shows the two-dimensional hexagonal Brillouin zone of the ${\mathrm{WSe}}_{2}$ monolayer 
with special points denoted by gray dots, and the blue (red) line that starts from the $\Gamma$ point indicates 
$\textbf{Q}$ along $\Gamma$M ($\Gamma$K) direction. Since our emphasis is on the excitons that are localized around the K and $\rm K'$ points, 
as shown in Fig. \ref{Fig:BZ}(b), we further shift these arrows to the K and $\rm K'$ points to illustrate 
the corresponding transitions for a given $\textbf{Q}$.
Figure \ref{Fig:exciton_band} illustrates the exciton band structure of the ${\mathrm{WSe}}_{2}$ monolayer obtained 
by solving SC-BSE for the $\textbf{Q}$ along both $\rm \Gamma M$ and $\rm \Gamma K$ directions. The exciton excitation 
energies are black dots, and the gray dashed lines serve as a guide to the eye for a better visualization of the exciton band structure. 
The excitation energy of the A exciton at $\textbf{Q} \rightarrow 0$ is indicated by the horizontal red line. 
The four lowest-energy exciton excitation energies correspond to the A exciton and its dark counterpart. 
We then present these four exciton bands with the fitting curves (blue lines). Following Ref.~\cite{Qiu2015} 
(also see supplementary note 2 in the SM~\cite{Stone-SM}, we fit the two lowest dark excitons 
and the lower band of the A exciton to
\begin{equation}
\begin{aligned}
\Omega(\textbf{Q})=\Omega(0) + A_{2}\left | \textbf{Q} \right |^2.
\end{aligned}
\label{Eq:parabolic_fit}
\end{equation}

The upper band of the A exciton clearly demonstrates a non-parabolic behavior. Consequently, we fit it to a third-order polynomial, defined as 
\begin{equation}
\begin{aligned}
\Omega(\textbf{Q})=\Omega(0) + A_{1}\left |\textbf{Q}\right | + A_{2}\left | \textbf{Q} \right |^2 + A_{3}\left | \textbf{Q} \right |^3.
\end{aligned}
\label{Eq:fitting}
\end{equation} 
We choose the direction of fitting to be $\Gamma$M, which has more data points than $\Gamma$K, and the fitting range spans 
$\rm \textbf{Q} = 0.15  \textup{~\AA}^{-1}$ (about 8 $\%$ of the size of Brillouin zone). Remarkably, the excitation energies 
in the $\Gamma$K direction align closely with the fitting curves shown in Fig. \ref{Fig:exciton_band}, suggesting isotropy in 
the excitation energies of the A exciton and its dark counterpart.

At $\textbf{Q}=0$, the two lowest bright excitons responsible for the A resonance in Fig. \ref{Fig:pol} are energetically degenerate 
with the excitation energy of $\Omega(0)$, as required by time-reversal symmetry \cite{Qiu2015,Sauer2021}. On the other hand, 
we find that the two lowest dark excitons of the ${\mathrm{WSe}}_{2}$ monolayer exhibit a small but nonzero energy splitting (4 meV), 
opposed to the ${\mathrm{MoS}}_{2}$ monolayer where the lowest two spin-forbidden dark excitons remain degenerate \cite{Qiu2015,Qiu2021}. 
This behavior is consistent with the previous theoretical study \cite{Deilmann2019}, which presents the splitting of roughly 12 meV. 
In supplementary note 3 ~\cite{Stone-SM}, we show that the dark exciton energy splitting converges with the increasing vacuum thickness, 
indicating it is not caused by spurious interlayer interactions in supercell periodic calculations. Nevertheless, the underlying reason remains unclear. 
There are two main factors resulting in the excitation energy difference between the bright A exciton and its dark counterpart 
(denoted as $\Delta_{DA}$). (i) The energy difference between the two lowest conduction bands is 42 meV (see Table \ref{table:QPband}). 
(ii) The lowest conduction band exhibits a heavier effective mass compared to the second lowest conduction band, which gives rise 
to a greater exciton binding energy of the dark excitons than the bright excitons. These factors add up to an excitation energy splitting 
of 76 meV, which agrees well with the finding of a previous theoretical study \cite{Deilmann2019} (see TABLE \ref{table:exciton mass}).

At finite $\textbf{Q}$, the degeneracy of the two bright excitons is lifted due to the electron-hole exchange interaction \cite{Qiu2015,Sauer2021}, and they develop into two branches. The lower bright exciton band shows a parabolic dispersion behavior. According to our polynomial fitting, we find the exciton effective mass of the lower band of the A exciton (denoted as $M_{B1}$) to be 0.857 $m_{0}$, where $m_{0}$ is free electron mass, and its value agrees well with the previous theoretical study (see Table \ref{table:exciton mass}). Note that $M_{B1}$ is significantly enhanced compared to the total mass of the free electron-hole pair $M = m_{v} + m_{c_{2}}$ (roughly 0.594 $m_{0}$), this is due to the electron-hole interaction, including the direct and exchange terms \cite{Qiu2015}. On the other hand, the upper band of the A exciton does not exhibit a clear parabolic dispersion behavior in the small $\textbf{Q}$ region (see Fig. \ref{Fig:exciton_band}). 
This non-parabolic behavior of the A exciton band is absent from the previous study by Deilmann $et\,al.$ \cite{Deilmann2019}. However, they adopted the 3D Coulomb potential in solving BSE without using the 2D truncated Coulomb interaction. In fact, according to the $\textbf{Q} \cdot \textbf{p}$ perturbation theory, and employing the fact that the two degenerate bright excitons are highly localized at the $\rm K$ and $\rm K'$ valleys, one can construct an effective BSE Hamiltonian in the tight-binding limit and diagonalize it to obtain the exciton band structure, as demonstrated by Qiu $et\,al.$ \cite{Qiu2015} and Sauer $et\,al.$ \cite{Sauer2021}. For this effective Hamiltonian the lowest bright exciton bands will undergo a split, leading to the formation of the non-parabolic upper band (labeled $\Omega_{l}(\textbf{Q})$) 
\begin{equation}
\begin{aligned}
\Omega_{l}(\textbf{Q})=\Omega(0) + 2A\left | \textbf{Q} \right | + (\frac{\hbar^2}{2M} + \alpha + \beta + |\beta^\prime|)\left | \textbf{Q} \right |^2
\end{aligned}
\label{Eq:nonanalytic}
\end{equation}
and the parabolic lower band (labeled $\Omega_{p}(\textbf{Q})$)
\begin{equation}
\begin{aligned}
\Omega_{p}(\textbf{Q})=\Omega(0) + (\frac{\hbar^2}{2M} + \alpha + \beta - |\beta^\prime|)\left | \textbf{Q} \right |^2
\end{aligned}
\label{Eq:parabolic}
\end{equation}
where $\Omega(0)$ is the exciton excitation energy at $\textbf{Q} \rightarrow 0$, $M = m_e + m_h$ is the total mass 
of the effective masses of the electron and hole that primarily constitute the exciton.
Here, $A$ is the proportionality constant of the linear terms in the intravalley and intervalley exchange i
interaction \cite{Qiu2015,Qiu2021}, $\alpha$ is the proportionality constant from the $Q^2$ contribution from the 
direct term, $\beta$ and $\beta^\prime$ are the $Q^2$ contribution from intravalley and intervalley exchange terms, respectively.

Due to the interplay of the long-range part (i.e. the $\textbf{G}=0$ term) of the intravalley and intervalley exchange 
interaction \cite{Qiu2015,Qiu2021}, the upper band of the A exciton has a linear term in \textbf{Q}, 
which is absent from the lower band. It is this linear term that yields the energy splitting 
between the two branches at small $\mathbf{Q}$. Note that the linear term is a distinct signature of a quasi-2D system 
(i.e., adopting the 2D Coulomb potential), and it causes the non-parabolic behavior of the upper branch, 
as seen in our {\it ab initio} calculations and other theoretical studies \cite{Qiu2015,Sauer2021,Qiu2021}.

The two lowest dark exciton bands, on the other hand, exhibit the parabolic dispersion behaviors, and the 4 meV energy separation observed at $\textbf{Q}=0$ persists at finite $\textbf{Q}$. Table \ref{table:exciton mass} summarizes the exciton effective masses extracted from the fitting curves. The exciton effective masses of the two lowest dark excitons (denoted as $M_{D1}$ and $M_{D2}$) are comparable, and they are significantly heavier than $M_{B1}$ due to the following two factors. (i) The total mass of the free electron-hole pairs that comprise the dark exciton $M = m_{v} + m_{c_{1}}$ is heavier than that of the bright exciton $M = m_{v} + m_{c_{2}}$ since $m_{c_{1}} > m_{c_{2}}$. (ii) The difference between the intravalley and intervalley exchange interaction will further decrease $M_{B1}$ \cite{Qiu2015}, while both terms are absent from the dark excitons.

\begin{table}[t]
\caption{The energy difference between the bright A exciton and its dark counterpart ($\Delta_{DA}$). The exciton effective mass 
of the parabolic lower band of the bright exciton ($M_{B1}$) and the two lowest dark excitons ($M_{D1}$ and $M_{D2}$) 
extracted from the fitting curves. The available theoretical values are listed.}
\begin{ruledtabular}
\begin{tabular}{l c c c c}
 & $\Delta_{DA}$ (meV) & $M_{B1}$ ($m_{0}$) & $M_{D1}$ ($m_{0}$) & $M_{D2}$ ($m_{0}$) \\
 \hline
 SC-BSE & 76  &  0.857 & 1.039 & 1.044 \\
 Theory & 70\footnote{Reference \cite{Deilmann2019}} & 0.877\footnote{Reference \cite{Sauer2021}} & ... & ... \\
\end{tabular}
\end{ruledtabular}
\label{table:exciton mass}
\end{table}

\subsection{Electron energy-loss spectrum}

\subsubsection{Momentum-dependent EEL spectrum}

\begin{figure}[tbph]
    \centering
    \includegraphics[width=8.5cm]{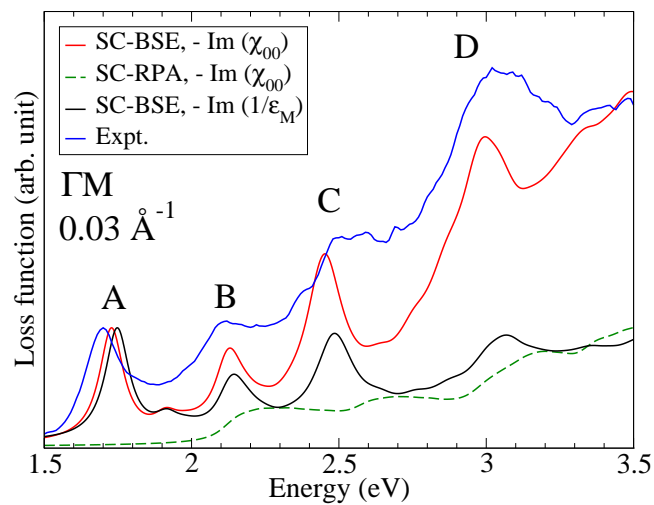}
    \caption{The calculated momentum-dependent EEL spectra of the ${\mathrm{WSe}}_{2}$ monolayer along $\Gamma$M direction for $Q = 0.03  \textup{~\AA}^{-1}$. Within the SC-RPA framework, we show the EEL spectrum obtained via Eq. (\ref{Eq:eels_chi00}) (green dashed line). Within the SC-BSE framework, we show the EEL spectra obtained via Eq. (\ref{Eq:eels_chi00}) (red solid line) and Eq. (\ref{Eq:eels_tradition}) (black solid line). The experimental EEL spectrum taken from Ref. \cite{JinhuaHong2020} (blue solid line) is also shown for comparison. All the theoretical spectra have been normalized based on the experimental A exciton peak intensities. The first four pronounced EELS peaks are labeled by A, B, C, and D.}
    \label{Fig:EELS_0_03}
\end{figure}

\begin{figure*}[tbph]
    \centering
    \includegraphics[width=17cm]{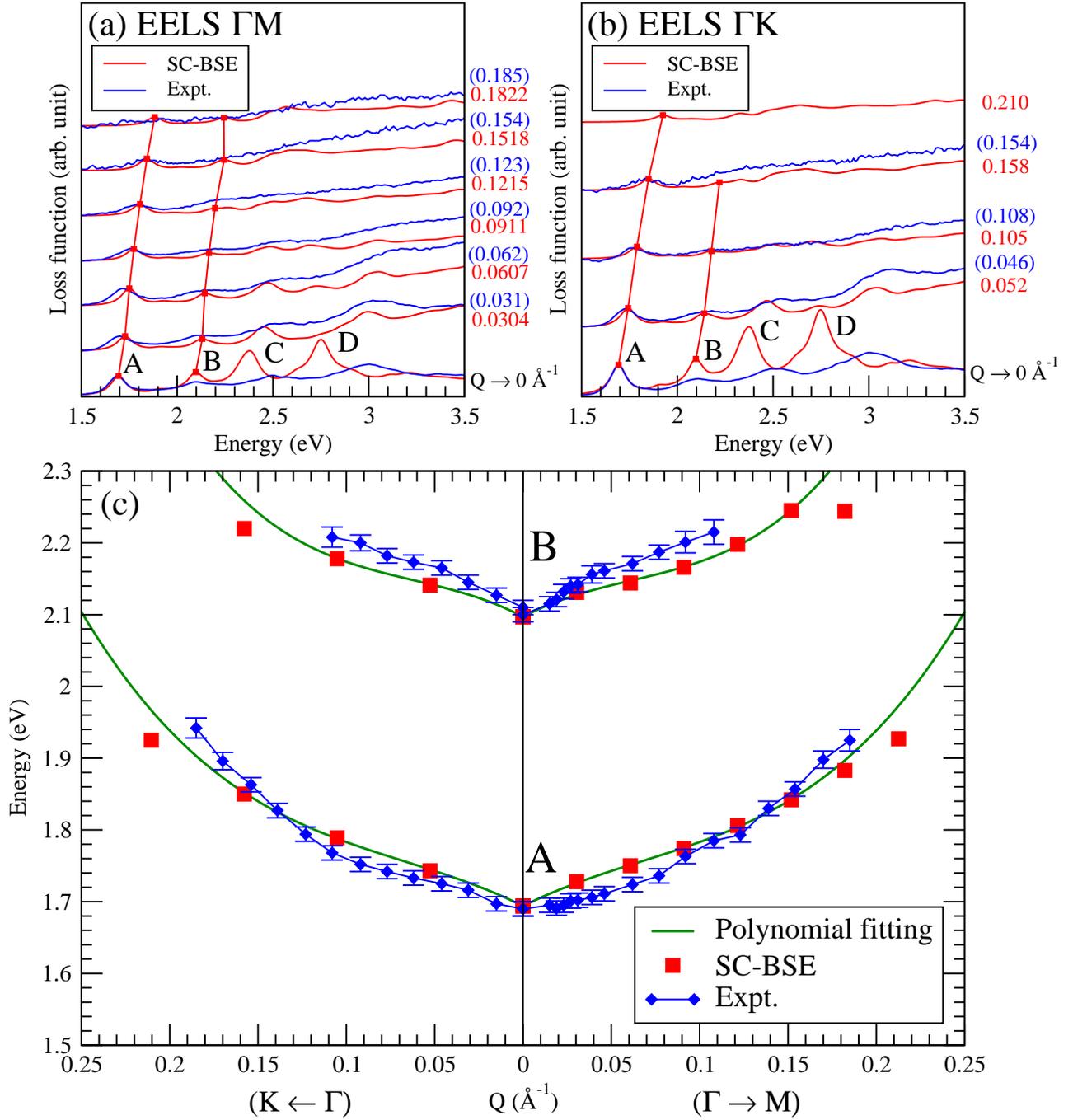}
    \caption{The momentum-dependent EEL spectra of the ${\mathrm{WSe}}_{2}$ monolayer along (a) $\Gamma$M and (b) $\Gamma$K 
directions obtained within the SC-BSE framework (red) and experimental measurement (blue) \cite{JinhuaHong2020}. The corresponding 
EELS peaks at $\textbf{Q} \rightarrow 0$ are labeled, and the theoretical A and B exciton peak dispersions are indicated 
by red squares connected with red lines, serving as a guild for clarity. The theoretical spectra have been normalized based 
on the experimental A exciton peak intensities. (c) The positions of the A and B EELS peaks along $\Gamma$M and $\Gamma$K directions 
(red squares). We simulate the A and B EELS peak dispersion curves by fitting in $\Gamma$M direction with 
a third-order polynomial (green line). The experimental EELS peak dispersion curves with error bars (blue diamonds) \cite{JinhuaHong2020} 
are also shown for comparison.}
    \label{Fig:eels_dispersion}
\end{figure*}

Figure \ref{Fig:EELS_0_03} shows the experimental and calculated momentum-dependent EEL spectra of the ${\mathrm{WSe}}_{2}$ monolayer 
along $\Gamma$M direction for $Q = 0.03  \textup{~\AA}^{-1}$. The experimental spectra is taken from Ref. \cite{JinhuaHong2020}. First, we consider the two spectra evaluated using Eq. (\ref{Eq:eels_chi00}) (\emph{Definition 2}) labelled $\mathrm{Im}(\chi_{00})$ in the figure. Within the scissor-corrected random phase approximation (SC-RPA) framework (green dashed line), the EEL spectrum fails to capture the main features of the experimental spectrum. This is because the strong excitonic effects dominate the spectrum in the low-energy and small $\textbf{Q}$ regimes presented in Fig. \ref{Fig:EELS_0_03}, and RPA framework omits the attractive electron-hole screened Coulomb interaction that is responsible for the formation of the excitons \cite{Onida2002}.
Remarkably, after including the excitonic effects by solving SC-BSE, the corresponding EEL spectrum (red solid line) captures the four distinct EELS peaks (labeled A, B, C, and D in Fig. \ref{Fig:EELS_0_03}), which agrees well with the experimental spectrum. Next, we consider the EEL spectrum that is determined using the inverse of the macroscopic dielectric function (Eq. (\ref{Eq:eels_tradition})) (\emph{Definition 1}) labelled $-\mathrm{Im}(1/\varepsilon_M)$ in the figure. Here the macroscopic dielectric function is calculated within the SC-BSE framework using Eq. (\ref{Eq:eps}) and a truncated Coulomb interaction, and we have used the effective thickness to be half of the cell thickness of bulk ${\mathrm{WSe}}_{2}$. The EEL spectrum calculated using Eq. (\ref{Eq:eels_tradition}) exhibits peaks at different positions and varying relative intensities compared to the EEL spectrum derived from Eq. (\ref{Eq:eels_chi00}). The latter demonstrates superior performance over the former when compared to the experimental spectrum. This sheds light on the importance of using the suitable formula of EEL spectrum, as we have discussed in Sec. \ref{Electron energy-loss spectrum}. 

Figures \ref{Fig:eels_dispersion}(a) and \ref{Fig:eels_dispersion}(b) illustrate the calculated EEL spectra of the ${\mathrm{WSe}}_{2}$ monolayer 
compared to the experimental spectra taken from Ref. \cite{JinhuaHong2020} along the $\Gamma$M and $\Gamma$K directions, respectively. 
At $\textbf{Q} \rightarrow 0$, the theoretical EEL spectrum of the ${\mathrm{WSe}}_{2}$ monolayer is the same as the optical absorbance 
(see Fig. S7(a)) as explained in Sec. \ref{Electron energy-loss spectrum}. Alternatively, this can be understood as the fact that 
the Coulomb interaction in reciprocal space goes as $1/Q$ for the small momentum transfers in a two-dimensional system, 
which effectively turns off the long-range coupling between electrons, thereby curbs the formation of the plasmon \cite{Nerl2017,JinhuaHong2022}. 
As shown in the SM~\cite{Stone-SM}, the EEL spectra are isotropic in the optical limit (see Fig. S7(b)). At $\textbf{Q} \rightarrow 0$, 
the theoretical EEL spectra can reproduce the two prominent peaks, labelled A and B, in the experimental spectra. However, 
the discrepancy between the theoretical and experimental spectra in the higher energy window is noticeable, which could stem 
from the fact that the zero momentum transfer can never be achieved in the experiments \cite{Abajo2010,JinhuaHong2020,JinhuaHong2021}, 
while we can directly set it to be zero in the theoretical calculations. This discrepancy is resolved for finite \textbf{Q} 
as shown in Fig. \ref{Fig:EELS_0_03}, where the theoretical spectra can accurately reproduce the experimental spectra.

We derive the EELS peak dispersion curves by monitoring the positions of peaks in the EEL spectra at different momentum transfers. 
Figure \ref{Fig:eels_dispersion}(c) shows the dispersion curves of the A and B peaks derived from the EEL spectra, 
alongside the experimental dispersion curves taken from Ref. \cite{JinhuaHong2020}. Like the exciton band structure, 
we fit the A and B EELS peak dispersion relations with a third-order polynomial (green line). We choose the direction 
of fitting to be $\Gamma$M, and the fitting range spans $Q = 0.15  \textup{~\AA}^{-1}$ (about 8 $\%$ 
of the size of Brillouin zone). Within the range of fitting, the third-order polynomial gives a good description of 
the dispersion curves for both A and B peaks, and the EELS peaks in the $\Gamma$K direction align closely with the fitting curves, 
indicating these peak dispersion curves are isotropic. 
Our A and B EELS dispersion curves agree well with the experimental measurements in both $\Gamma$M and $\Gamma$K directions. 
Interestingly, as shown in Fig. \ref{Fig:exciton_band}, our A EELS dispersion curve follows the upper band of the A exciton and 
thus exhibits a non-parabolic behavior. However, it is concluded in Ref. \cite{JinhuaHong2020} that the A EELS dispersion curves 
are isotropic and parabolic in both $\Gamma$M and $\Gamma$K directions. Despite this, we note that the experimental A EELS dispersion 
can be interpreted as either parabolic or non-parabolic within the error bars of the experiment \cite{JinhuaHong2020}. 
Therefore, the qualitative behavior of the A EELS dispersion curve remains an open question. On the other hand, both experimental 
and theoretical B EELS dispersion curves clearly manifest non-parabolic behavior. In the next section,  we will explain why 
the EELS peaks should follow the upper band of the A exciton and thus show the non-parabolic dispersion behavior. 

\subsubsection{Selective excitation of excitons via EELS}

To find the origin of the non-parabolic dispersion of the A exciton peaks, we show in Fig. \ref{Fig:exciton_band} 
the dispersion curves of the A EELS peaks obtained within the SC-BSE framework (red squares) on top of the exciton band 
structure (black dots). Although numerous exciton states are within the energy range depicted in Fig. \ref{Fig:exciton_band}, 
not all of them contribute to the EEL spectrum. For example, the two lowest-energy dark excitons are not dipole-allowed 
and are therefore not visible in the EEL spectrum. Surprisingly, we find the calculated A EELS peaks unambiguously track 
the non-parabolic upper band of the A exciton, whereas the parabolic lower band does not contribute to the EEL spectrum. 
To understand the principles governing the electronic excitation mechanism in electron energy-loss spectroscopy, 
we analyze the definition of the EEL spectrum in Eq. (\ref{Eq:eels_chi00}). Based on first-order $\textbf{Q} \cdot \textbf{p}$ 
perturbation theory \cite{Qiu2015,Qiu2021}, and keeping terms up to the lowest order in $\textbf{Q}$, the contribution 
from an exciton state $|{\lambda}\rangle$ to the EEL spectrum is
\begin{equation}
\begin{aligned}
& \left |
\sum\limits_{vc\textbf{k}}A^{\lambda,\textbf{Q}}_{vc\textbf{k}}\rho_{vc\textbf{k}}(\textbf{Q},0) \right |^2\approx \\
& \left |\textbf{Q} \cdot \frac{\hbar}{m} 
\sum\limits_{vc\textbf{k}}A^{\lambda,0}_{vc\textbf{k}}\frac{\langle{u_{v\textbf{k}}|\textbf{p}|u_{c\textbf{k}}}\rangle}{E^{\mathrm{QP}}_{v\textbf{k}}-E^{\mathrm{QP}}_{c\textbf{k}}} \right |^2 = \left | \frac{\hbar}{m}\textbf{Q} \cdot \textbf{P}_{\lambda} \right |^2 .
\end{aligned}
\label{Eq:eels_weight}
\end{equation}
That is, the contribution of an exciton to the EEL spectrum is directly related to the inner product of its center-of-mass momentum 
and dipole matrix element of the exciton. Since our focus is primary on the in-plane momentum transfers 
(i.e., $\mathbf{Q}=\mathbf{Q}_{||}$), this indicates that the electron-energy-loss spectroscopy will selectively 
probe longitudinal excitons, while transverse excitons will have negligible contribution to the EEL spectrum. 
Remarkably, the non-parabolic upper band of the A exciton consists of longitudinal excitons, while the parabolic lower 
band of the A exciton comprises transverse excitons \cite{Denisov1973,Sauer2021}. Therefore, the A EELS peaks derived 
from the EEL spectrum will follow the non-parabolic upper band of the A exciton. Same argument can be applied to the dispersion
of the B EELS peaks because the B exciton bands will also split into the non-parabolic upper band and the parabolic 
lower band at finite $\textbf{Q}$ \cite{Sauer2021}, with the upper (lower) band consists of the longitudinal (transverse) 
excitons \cite{Sauer2021}. We note that this argument is not exclusive to the ${\mathrm{WSe}}_{2}$ monolayer. 
Instead, it can be considered the signature of quasi-2D materials. 

Next, we provide an alternative perspective on this phenomenon. The EELS techniques exploit swift electrons that interact with a specimen to unravel its underlying excitation, such as plasmon and exciton \cite{Abajo2010,JinhuaHong2020}. Considering an electron moving along the $z$ axis with velocity $v$, here $z$ is the direction perpendicular to the quasi-2D sheet, and the total momentum transfer of the electron $\mathbf{Q}$ consists of its in-plane component $\mathbf{Q}_{||}$ and out-of-plane component $q_z\mathbf{\hat{z}}$. The electric field produced by it in vacuum is \cite{Abajo2010}
\begin{equation}
\begin{aligned}
\textbf{E}(\mathbf{r}, \omega) = \frac{ie}{\pi} \int d^{3}\mathbf{Q} 
\frac{\mathbf{Q}-\frac{\omega}{c^2}\mathbf{v}}{\left | \mathbf{Q} \right |^2-\frac{\omega^2}{c^2}}
e^{i\mathbf{Q}\cdot \mathbf{r}}
\delta(\omega-\mathbf{Q}\cdot \mathbf{v}). 
\end{aligned}
\label{Eq:Electric_field}
\end{equation}
By performing the in-plane Fourier transform on the electric field, the Fourier coefficient of the electric field 
can be decomposed into the in-plane component
\begin{equation}
\begin{aligned}
E_{||}(\textbf{Q}_{||})=\frac{ie}{\pi}
\frac{e^{i\frac{\omega}{v}z}}{\left | \mathbf{\mathbf{Q}_{||}} \right |^2+\frac{\omega^2}{v^2}(1-\frac{v^2}{c^2})}\mathbf{\mathbf{Q}_{||}} 
\end{aligned}
\label{Eq:Electric_field_in_plane}
\end{equation}
and the out-of-plane component
\begin{equation}
\begin{aligned}
E_z(\textbf{Q}_{||})=\frac{ie}{\pi}
\frac{e^{i\frac{\omega}{v}z}}{\left | \mathbf{\mathbf{Q}_{||}} \right |^2+\frac{\omega^2}{v^2}(1-\frac{v^2}{c^2})}(1-\frac{v^2}{c^2})q_{z}\mathbf{\hat{z}}.
\end{aligned}
\label{Eq:Electric_field_out_of_plane}
\end{equation}
We derive Eq. (\ref{Eq:Electric_field_in_plane}) and Eq. (\ref{Eq:Electric_field_out_of_plane}) in supplementary note 4~\cite{Stone-SM}. 
Interestingly, the in-plane component and out-of-plane component are proportional and parallel to $\mathbf{Q}_{||}$ and $q_z\mathbf{\widehat{z}}$, 
respectively. 

In the previous experimental setup ~\cite{JinhuaHong2020}, the incident electron was moving in the out-of-plane ($z$) direction. 
Due to the high velocity of the incident electrons, the magnitude of $q_z$ is considerably smaller than that
of $\mathbf{Q}_{||}$ \cite{Abajo2010,JinhuaHong2020}. That is, the electric field primarily lies in-plane, 
which is parallel to $\mathbf{Q}_{||}$. The electric field then probe an exciton with its exciton dipole matrix element 
in the direction of $\mathbf{Q}_{||}$, which is exactly the center-of-mass momentum of this exciton. 
This coincides with the definition of a longitudinal exciton, explaining that the EELS
will only selectively probe the longitudinal exciton consisting of the upper band of the A and B exciton 
under this experimental setup.

It will also be interesting to probe the other exciton bands, such as the lower parabolic band of the A and B 
exciton using EELS techniques. Given that the lower parabolic band of the A and B exciton consists of the 
transverse exciton with its exciton dipole matrix element pointing to the $z$ direction \cite{Sauer2021}, 
we consider the electric field from incident electrons at a specific momentum transfer also pointing primarily 
in the $z$ direction, which can be achieved by rotating the direction of the incident electron beam to the $x$ axis. 
In this case, the Fourier component of the electric field in the $x$ direction will be small compared to its $y$ component
\begin{equation}
\begin{aligned}
E_{y}(Q_{y},Q_{z})=\frac{ie}{\pi}
\frac{e^{i\frac{\omega}{v}x}}{(Q_{y}^{2}+Q_{z}^{2})+\frac{\omega^2}{v^2}(1-\frac{v^2}{c^2})}
Q_{y}
\end{aligned}
\label{Eq:Electric_field_y}
\end{equation}
and the $z$ component
\begin{equation}
\begin{aligned}
E_{z}(Q_{y},Q_{z})=\frac{ie}{\pi}
\frac{e^{i\frac{\omega}{v}x}}{(Q_{y}^{2}+Q_{z}^{2})+\frac{\omega^2}{v^2}(1-\frac{v^2}{c^2})}
Q_{z}.
\end{aligned}
\label{Eq:Electric_field_z}
\end{equation}
By choosing a momentum transfer with $Q_{z} > Q_{y} \gg q_{x} = \omega/v$, the electric field will lie primarily in 
the $z$ direction to couple with the exciton polarization, while the motion of the exciton is constrained to the $x-y$ plane 
due to the monolayer structure. Therefore, we can probe the lower band of the A and B exciton consisting of the transverse exciton 
in this experimental setup. However, it should be noted that the derived formula of Eq. (25) is only valid 
for the transmission EELS with the direction of the incident electron beam along the $z$-axis.

\section{CONCLUSIONS}
In summary, we have carefully studied the different EEL spectrum formulas for quasi-2D materials that have been widely 
used in the literature, and established an appropriate definition of the EEL spectrum, a formula adopted 
in many previous studies yet lacking a clear explanation until now. 
By utilizing the state-of-the art {\it ab initio} BSE calculations with the relativistic spin-orbit coupling included,
we have investigated the excitons with finite center-of-mass momentum 
(i.e., exciton band structure) and, with the proper EELS formula mentioned above, have also calculated 
the EEL spectrum of the ${\mathrm{WSe}}_{2}$ monolayer with a perpendicular incident electron beam,
as in the previous experimental setup~\cite{JinhuaHong2020}.
We identify five exciton states through the calculated optical absorbance of the ${\mathrm{WSe}}_{2}$ monolayer, 
including the lowest-energy bright A exciton.
Our theoretical analysis reveals that due to the electron-hole exchange interaction, the bright A exciton bands 
which are degenerate at the zero momentum transfer $Q=0$, split at finite momentum transfers. 
The lower branch of the A exciton and also the dark exciton bands exhibit parabolic characteristics, 
while the the upper band of the A exciton shows a non-parabolic dispersion relation. 
Interestingly, the dispersion of the calculated lowest-energy EELS peaks follows closely the non-parabolic 
upper band of the A exciton, and also agrees well with the experimental spectrum in the low-energy-loss regime~\cite{JinhuaHong2020}.
Furthermore, our analysis also shows that EELS experiments would selectively probe those exciton bands
having their dipole being parallel to the transfered momentum. 
This explains why only the upper band of the A exciton, which is a longitudinal exciton with
an in-plane dipole moment, was observed in the previous experiment~\cite{JinhuaHong2020}.
Our findings thus suggest further EELS experiments with different orientations of the incident electron beam 
to measure other branches, such as the parabolic lower band of the A exciton, of the exciton band structure 
and hence to gain a full understanding of the exciton dynamics in quasi-2D materials.

\section{Acknowledgments}
The authors would like to express their gratitude to the authors of \cite{JinhuaHong2020} for generously sharing 
their digital data of the experimental EEL spectra with us. Y.-C.S and G.-Y.G thank the support from the National 
Science and Technology Council and National Center for Theoretical Sciences (NCTS), Taiwan.
F.N has received funding from the European Union’s Horizon 2020 research and innovation program under 
the Marie Skłodowska-Curie grant agreement No. 899987.  (EuroTechPostdoc2). Fruitful discussions with K. S. Thygesen, 
M. Ohm Sauer and T. G. Pedersen are also gratefully acknowledged.

{}	

\end{document}